# Nonlinear Image Formation by Optical Superlattices

Bo Yang,[1,2] Xu-Hao Hong,[1,3] Rong-Er Lu,[1,2] Yang-Yang Yue,[1,2] Chao Zhang,[1,2,*]
Yi-Qiang Qin,[1,2,†] and Yong-Yuan Zhu[1,3]

[1] *National Laboratory of Solid State Microstructures and Collaborative Innovation Center of Advanced Microstructures and Key Laboratory of Modern Acoustics, Nanjing University, Nanjing 210093, China*
[2] *College of Engineering and Applied Sciences, Nanjing University, Nanjing 210093, China*
[3] *School of Physics, Nanjing University, Nanjing 210093, China*
*Corresponding author: zhch@nju.edu.cn*
[†] *Corresponding author: yqqin@nju.edu.cn*

**ABSTRACT**

The angular spectrum theory is applied to the nonlinear harmonic generation process in optical superlattices. Several explicit and analytical structure functions are deduced to design optical superlattices for various purposes. Employing this method, nonlinear image formation is achieved during the second-harmonic generation process in a properly designed optical superlattice. This method is universal for both nonlinear beam shaping and nonlinear arbitrary image formation. The theory has been experimentally validated in two-dimensional optical superlattice of $LiTaO_3$ crystals and the results agree well with the theoretical prediction. This work not only extends the application of optical superlattices, but also opens a new area for imaging technologies.

Image formation, such as display, microscopy, photography, and holography, has been an active research topic in optics for a long time [1-5]. In the past decades, the study of metamaterials and metasurface materials discovered many novel phenomenon such as negative refraction and optical cloaking, which upgraded the research of image formation to a new level [6-12]. Nonlinear optics can be also introduced to the study of image formation. Nonlinear Abbe theory, generalized in a straightforward manner to include spatial nonlinearity, has been proved to be useful for increasing the resolution ability of imaging [13]. More recently, a few researchers have paid attention to the image formation in nonlinear optical superlattice (OSL). Nonlinear Talbot Effect, by investigating near-field self-imaging effect of the second harmonic (SH), provides a new way to observe domain structures of the OSLs [14, 15]. Nonlinear beam shaping is another hot topic in nonlinear optics. Nonlinear generation of Airy beam and vortex beam has been realized by many different methods [16-22]. A common used method to design the domain structure for beam shaping is introducing a transverse dependent phase in a periodic structure [23]. Nonlinear volume holography is another effective method for the same purpose, where the diffraction effect is fully considered [22].

Technically, the nonlinear image formation can be also treated as a beam shaping process, that is,

shaping the fundamental wave (FW) from plane wave into an arbitrary given SH image. Therefore, the methods used for OSLs design of nonlinear beam shaping should be useful for nonlinear image formation. Yet, we found that it will encounter difficulties when applying these methods to a general imaging formation process.

The nonlinear processes in two-dimensional (2D) OSLs can be usually classified into two different configurations. One is that the polarization direction is perpendicular to the propagation direction, such as the nonlinear generation of Airy beams [16, 22], and the focused second-harmonic generation (SHG) using LQPM (local quasi-phase-matching) [24, 25]. The other configuration is that the polarization direction is along with the propagation direction, such as the generation of nonlinear vortex beams [17, 18] and the nonlinear Talbot effect [15]. These two configurations are usually regarded as beam shaping and image formation respectively. Is there any method can be used to design OSLs for both of the two configurations? The answer is positive. The first method we should consider is the nonlinear volume holography, since holography is a popular technique for linear image formation [26-29], which has also been applied to metasurface materials recently [30-33]. However, to design an OSL for nonlinear image formation with this method, the distribution of nonlinear harmonic wave in the OSLs must be known in advance. This is possible for Airy beams and Bessel beams where the wave functions for the beams are well known but not so easy for an arbitrary image. In general, to find out the SH distribution in the OSLs with a given image is an inverse problem [34]. It is well known that angular spectrum theory is a rigorous and efficient way to calculate the field distribution in free space with the knowledge of the field distribution at a given plane, which plays an important role in the linear imaging and wave propagation processes [35-38]. In this letter, we generalize the angular spectrum theory from linear optics to nonlinear optics and find that this method is suitable to solve the inverse problem in the nonlinear image formation process. With this theory, we realize arbitrary nonlinear image formation during the SHG process. Experiments as well as numerical simulations have been performed to verify the theory.

In a SHG process, the SH secondary source excited at plane $\Omega$, as shown in Fig. 1, satisfies a brief expression $dE_2(\alpha,\beta,\gamma) = -iKf(\alpha,\beta,\gamma)E_1^2(\alpha,\beta,\gamma)d\alpha$, here $E_1$, $E_2$ and $k_1$, $k_2$ are the electric field and wave-vector of the FW and the SH wave (SHW) respectively; $\alpha,\beta,\gamma$ represent the three-dimensional coordinates of OSLs; $K$ is a constant related to the nonlinear coefficient; $f(\alpha,\beta,\gamma)$ represents the domain structure function. To employ the angular spectrum theory, we assume that the angular spectrum of SHW at plane $\Omega$ is $a_{2\Omega}(k_{2\beta},k_{2\gamma})$, $k_{2\beta},k_{2\gamma}$ are the wave-factor components of SHW along the $\beta$, $\gamma$ axis and the propagation direction is along the $\alpha$ axis, so the SHW excited at plane $\Omega$ can be expressed as,

$$E_{2\Omega} = \int\int_{-\infty}^{\infty} a_{2\Omega}(k_{2\beta},k_{2\gamma})e^{i(k_{2\beta}\beta+k_{2\gamma}\gamma)}dk_{2\beta}dk_{2\gamma} \tag{1}$$

According to the angular spectrum theory, the wavefronts on two different planes are connected by the transfer function. Resulting from the linear propagation system after exciting, the transfer function between the plane $\Omega$ and the screen $\Sigma$ is $H = \exp[ik_{2a}(L-\alpha)]$ (where $k_{2a} = \sqrt{k_2^2 - k_{2\beta}^2 - k_{2\gamma}^2}$). An ideal SH image can be formed only when the angular spectrum of the SH wave at plane $\Sigma$ is the same as the spatial frequency of the objective image. Here we use $g(k_{2\beta},k_{2\gamma})$ to represent the spatial frequency of the image, and we have

$$g(k_{2\beta},k_{2\gamma}) = a_{2\Omega}(k_{2\beta},k_{2\gamma})H \tag{2}$$

With the above analysis, the relationship between the angular spectrum of secondary source and the spatial frequencies of the image can be established. Setting the FW to be a plane wave $E_1(\alpha,\beta,\gamma) = A_{10}(\beta,\gamma)\exp(ik_1\alpha)$, here $A_{10}(\beta,\gamma)$ is the initial amplitude, the structure function of the OSLs is determined by,

$$f(\alpha,\beta,\gamma) = sign\{-\int\int_{-k_2}^{k_2} g(k_{2\beta},k_{2\gamma})\exp[ik_{2\alpha}(\alpha-L)+i(k_{2\beta}\beta+k_{2\gamma}\gamma)-2ik_1\alpha]dk_{2\beta}dk_{2\gamma}\} \tag{3}$$

Eq. (3) is a domain structure function of nonlinear OSLs in general case. It can be degenerated into suitable forms to deal with the two different configurations we mentioned before.

For the first configuration, assuming the polarization direction is along the z axis and the propagation direction is along the x axis, the structure function becomes,

$$f(x,y) = sign\{-\int_{-k_2}^{k_2} g(k_{2y})\exp[ik_{2x}(x-L)+ik_{2y}y-2ik_1x]dk_{2y}\} \tag{4}$$

Here the x-component of the wave-vector satisfies $k_{2x} = \sqrt{k_2^2 - k_{2y}^2}$. The OSL structure defined by the above function is not only capable of forming an (one-dimensional) image, but also ensures the nonlinear process to meet the phase matching condition, as shown in Fig. 2.

Fig. 2(a) shows the OSL structure for the a dual-focused beam of SH wave with the same intensity. The SHW distribution is shown in Fig. 2(b). The OSL structure is obtained from Eq. (4) directly, where $g(k_{2y}) = \exp(-ik_{2y}y_1) + \exp(-ik_{2y}y_2)$, $y_1$, $y_2$ are the coordinates of the two focuses. Obviously, the structure is the superposition of two ellipse-like structures, which are basically the same as the one obtained by the LQPM technique[25]. Since the nonlinear beam shaping has been widely studied in the past several years, here we just exhibit a simple example for demonstration. The beam shaping of more complicated situation such as Airy beams can be realized as well, the resulting structure coincide well with that obtained by Holographic method.

For the second configuration, the nonlinear angular spectrum method is more universal. Here both the polarization direction and the propagation direction are along the z axis. The domain structure function can be determined by

$$f(x, y; z_0) = sign\{-\int\int_{-k_2}^{k_2} g(k_{2x}, k_{2y}) \exp[ik_{2z}(z_0 - L) + i(k_{2x}x + k_{2y}y) - 2ik_1 z_0] dk_{2x} dk_{2y}\} \quad (5)$$

The wave-vector component $k_{2z}$ satisfies $k_{2z} = \sqrt{k_2^2 - k_{2x}^2 - k_{2y}^2}$, and $z_0$ is the coordinate of the structure function plane. In this situation, a 2D nonlinear image can be obtained with a properly designed 2D OSL. Next, we will exhibit the numerical and experimental results of nonlinear imaging of an alphabet S and the badge of 'batman'.

The experimental setup is shown in Fig. 3, where a femtosecond mode-locked Ti:sapphire laser operating at a wavelength of 900nm and 600mW is chosen as the FW source. The structure is determined according to the structure function Eq. (5) with the given SH images and fabricated by the pulsed electric-field poling technique at room temperature [39]. The thickness of the sample slice is 500μm, and the poled area is 500μm×500μm as shown in Fig. 4(c) and (g). The distance between the imaging plane and the sample is set to be 5000μm. The FW laser beam from the Ti:sapphire laser is first shaped by a duplet lens system to obtain an almost parallel beam with a spot size of ~300μm and the polarization is parallel to the x axis of the sample. A long-pass filter is placed before the sample to filter the visible noise, and a short-pass filter is placed between the sample and the objective lens to eliminate the remaining FW. The SH patterns at different imaging planes are recorded by a microscope using CCD camera, which is able to move freely along the z axis controlled by a precision translation stage.

Fig. 4 shows the simulating and experimental results of nonlinear image formation. The OSL structures are shown in Fig. 4(a) and (e). In the experiment, we employ z-cut 2D OSLs of $LiTaO_3$ with both x-surfaces polished. Fig. 4(c) and (g) show the optical microscopic image of the sample surface after HF acid etching. The duty cycles of the OSLs are different with each other, and rely on the object. The scale of original images (object) is set to be the same with the poled area of OSLs which is 500μm×500μm. The distinctive SH images of the objects obtained by the numerical simulation with a finite difference method [40] are presented in Fig. 4(b) and (f) and the original images are shown in the insets in Fig. 4(b) and (f). The SH patterns of alphabet S and the badge of 'batman' recorded by CCD are exhibited in Fig. 4(d) and (h) respectively. The distance between the imaging plane and the OSL is ~5300μm, which is slightly deviated from the theoretical expectation. And there are some diffraction stripes around the main image. The reason for this may be the thickness of the OSL in the propagation direction cannot be neglected.

The image resolution is related to the angular spectrum. That is, the more higher-order angular spectrum is involved, the more details of the image can be revealed. Using a proper designed 3D OSL [41] if possible, the nonlinear imaging process will be phase matched and will supply more accurate angular spectrum. Unfortunately, the poling method for 3D OSLs is not feasible at present. Up to now, the 2D nonlinear OSL crystals are capable of performing different optical functions such as direct third-harmonic generation [39] and nonlinear Talbot effect[15] where the OSL structure is periodic or quasiperiodic. For more complicated functions more complicated OSL design methods are required. For example, for nonlinear harmonic focusing, the LQPM theory can be utilized [25]. For the beam shaping of nonlinear special beams such as Airy beams and Bessel beams, the nonlinear volume holography can be utilized [22]. For the latter case, an explicit analytical expression for the object wave is needed to determine the OSL structure. However, for an arbitrary image formation, there is no such an expression for the object wave. To find out the wavefront distribution for OSLs design is an inverse problem, and is not convenient using those two methods. We generalize the angular spectrum theory to nonlinear optics to solve the inverse problem for OSLs design. With this theory, we could design an OSL for generating arbitrary SH image at a given plane. The experimental results agree well with the expectation. This method is universal, it can be used in nonlinear image generation as well as in nonlinear beam shaping, and it can be also applied for the designing of other nonlinear photonic device for such as the biological detection and imaging.

This work is supported by the State Key Program for Basic Research of China (Grant No. 2012CB921502), the National Natural Science Foundation of China (Grant Nos. 11274163, 11274164, 11374150, 11504166 and 11574146), and Priority Academic Program Development of Jiangsu Higher Education Institutions of China (PAPD). The authors also thank Dr. D. M. Liu for the help in the experiment.

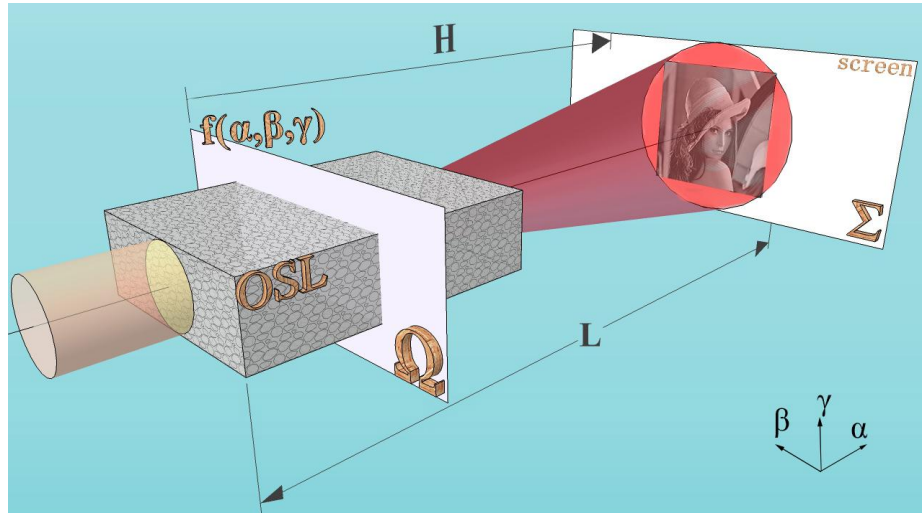

Fig. 1. Diagram of a nonlinear image formation process. $\Omega$ is a plane in optical superlattice vertical to the propagation direction, and the image plane is $\Sigma$. $f(\alpha,\beta,\gamma)$ represents the domain structure function. H is the transfer function.

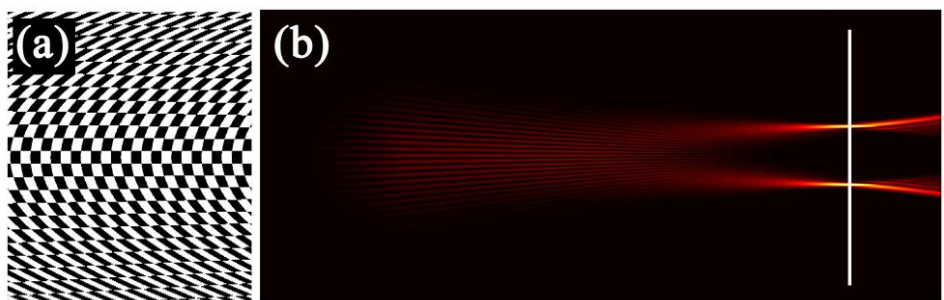

Fig. 2. Numerical simulation result of one-dimensional image formation using a 2D optical superlattice. (a) The structures of the OSLs for the dual-focused beam of SH wave. (b) The energy distribution of propagating SH beams.

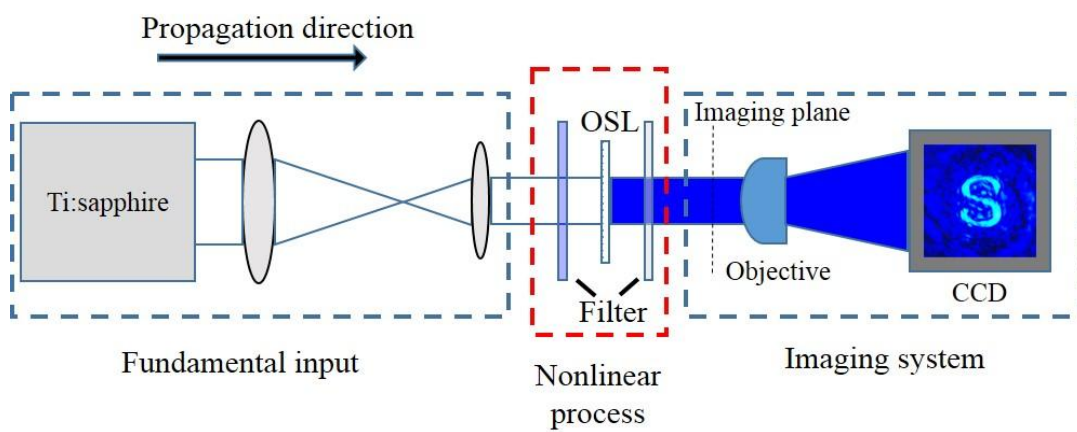

Fig. 3 Schematic of the experimental setup. The laser beam generated by the Ti:sapphire laser is shaped to be an almost parallel beam. The OSL is settled at the waist plane of the fundamental beam. The SH patterns are recorded with a CCD camera in the imaging system.

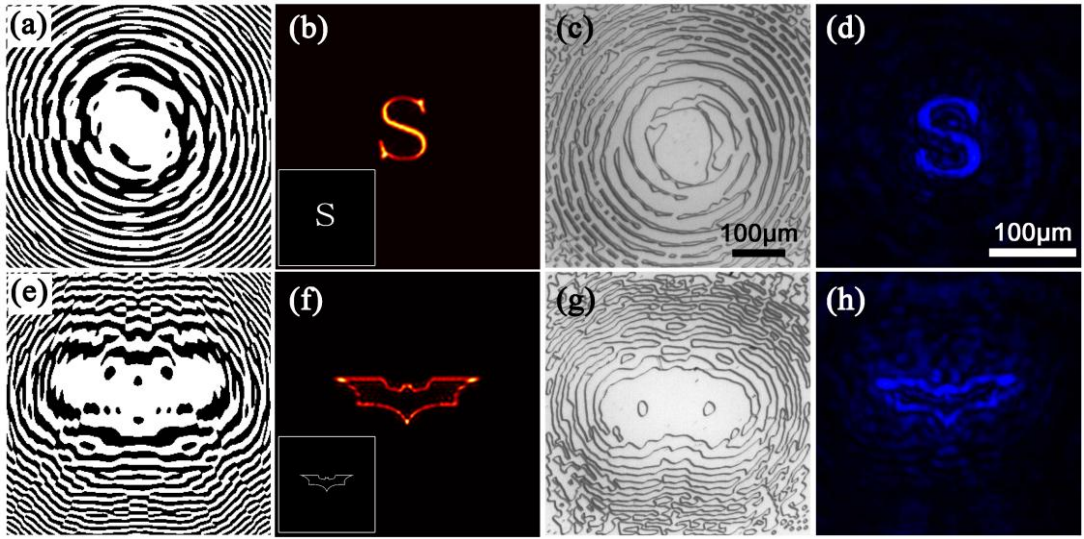

Fig. 4 Simulating and experimental results of nonlinear image formation. The original objects are inset in (b) and (f). (a), (b), (c), (d) and (e), (f), (g), (h) exhibit the simulated OSLs structure, numerical imaging results, optical microscopic images of OSLs, and the experimental imaging results of and alphabet S and the badge of 'batman' respectively.